\documentclass[reprint,superscriptaddress,amsmath,amssymb,aps,prd,twocolumn,floatfix,nofootinbib]{revtex4-1}
\usepackage{amsmath, amssymb}
\usepackage{listings}
\usepackage{graphicx}
\usepackage{dcolumn}
\usepackage{bm}
\usepackage{float,color}
\usepackage{xcolor}
\usepackage[normalem]{ulem}
\usepackage{tabularx}
\usepackage{mathtools} 
\usepackage{multirow}
\usepackage{amssymb}
\usepackage{scrextend}
\usepackage{hyperref}

\definecolor{teal}{rgb}{0., 0.5, 0.5}

\begin{document}

\title{Predictions of a simple parametric model of hierarchical black hole mergers}

\author{Parthapratim Mahapatra}\email{ppmp75@cmi.ac.in}
\affiliation{Chennai Mathematical Institute, Siruseri, 603103, India}
\author{Debatri Chattopadhyay}
\affiliation{School of Physics and Astronomy, Cardiff University, Cardiff, CF24 3AA, United Kingdom}
\author{Anuradha Gupta}
\affiliation{Department of Physics and Astronomy, The University of Mississippi, University, Mississippi 38677, USA}
\author{Marc Favata}
\affiliation{Department of Physics \& Astronomy, Montclair State University, 1 Normal Avenue, Montclair, New Jersey 07043, USA}
\author{B. S. Sathyaprakash}
\affiliation{Institute for Gravitation and the Cosmos, Department of Physics, Penn State University, University Park, Pennsylvania 16802, USA}
\affiliation{Department of Astronomy and Astrophysics, Penn State University, University Park, Pennsylvania 16802, USA}
\affiliation{School of Physics and Astronomy, Cardiff University, Cardiff, CF24 3AA, United Kingdom}
\author{K. G. Arun}
\affiliation{Chennai Mathematical Institute, Siruseri, 603103, India}
\affiliation{Department of Astronomy and Astrophysics, Penn State University, University Park, Pennsylvania 16802, USA}

\date{\today}

\begin{abstract}
The production of black holes with masses between $\sim 50M_{\odot}\hbox{--}130M_{\odot}$ is believed to be prohibited by stellar processes due to (pulsational) pair-instability supernovae. Hierarchical mergers of black holes in dense star clusters are proposed as a mechanism to explain the observations of binary black holes with component masses in this range by LIGO/Virgo.
We study the efficiency with which hierarchical mergers can produce higher and higher masses using a simple model of the forward evolution of binary black hole populations in gravitationally bound systems like stellar clusters. The model relies on pairing probability and initial mass functions for the black hole population, along with numerical relativity fitting formulas for the mass, spin, and kick speed of the merger remnant. We carry out an extensive comparison of the predictions of our model with {\tt clusterBHBdynamics} ({\tt cBHBd}) model, a fast method for the evolution of star clusters and black holes therein. For this comparison, we consider three different pairing functions of black holes and consider simulations from high- and low-metallicity cluster environments from {\tt cBHBd}. We find good agreements between our model and the {\tt cBHBd} results when the pairing probability of binaries depends on both total mass and mass ratio. We also assess the efficiency of hierarchical mergers as a function of merger generation and derive the mass distribution of black holes using our model. We find that the multimodal features in the observed binary black hole mass spectrum---revealed by the nonparametric population models---can be interpreted by invoking the hierarchical merger scenario in dense, metal-rich, stellar environments. Further, the two subdominant peaks in the GWTC-3 component mass spectrum are consistent with second and third-generation mergers in metal-rich, dense environments.
With more binary black hole detections, our model could be used to infer the black hole initial mass function and pairing probability exponents.  
\end{abstract}

\maketitle



\section{Introduction} \label{sec:intro}
Understanding the formation channels of binary black holes (BBHs) is one of the primary science goals of gravitational wave (GW) astronomy. The proposed formation scenarios of BBHs are broadly divided into two categories: isolated field binary evolution \citep{Dominik:2014yma, Belczynski:2016obo,Woosley:2016nnw,Marchant:2016wow,deMink:2016vkw} and dynamical binary formation in dense stellar environments~\citep{PortegiesZwart:1999nm,Miller:2008yw,Downing:2009ag,Rodriguez:2015oxa,Hurley2016,OLeary:2016ayz,Rodriguez:2016kxx,Rodriguez:2018pss,Petrovich:2017otm,Sedda2018,Fragione:2018vty,Zevin:2018kzq, Chattopadhyay:2022buz}. Among the $\sim$90 BBHs reported in the third LIGO-Virgo-KAGRA (LVK) gravitational-wave transient catalog (GWTC-3) \citep{GWTC-3}, a few binaries~\footnote{There are 8 (3) compact binaries in the GWTC-3 for which the mass posteriors of at least one of the companions exceeds $50M_\odot$ ($60M_\odot$) at 90\% credibility.} have at least one component's mass in the ``upper mass gap" (between $\sim 50 \mbox{--}130 M_{\odot}$). It is widely believed that this region is forbidden by stellar evolution due to pair-instability or pulsational pair-instability supernovae~\citep{Fowler1964, Barkat1967, Woosley:2016hmi, Farmer:2019jed, Farmer:2020xne, Renzo:2020lwl, Marchant:2018kun} and the result of fundamental physics at play in the evolution of massive stars \citep{Fryer:2000my}.

An alternate pathway to populate the upper mass gap is hierarchical assembly via multiple generations of BBH mergers \citep{Fragione:2017blf, Antonini:2018auk,Fragione:2020nib,Mapelli:2021syv, Britt:2021dtg, Banerjee:2017mgr, Chattopadhyay:2023arXiv}. In this scenario progenitor black holes (BHs) do not originate directly from the collapse of massive stars, but rather from the remnants of previous generations of BBH mergers. \citealt{Kimball:2020qyd} found evidence for hierarchical BBH mergers in the GWTC-2.
Because BBH mergers generically result in GW recoil~\citep{Fitchett83,Favata:2004wz},  this requires an astrophysical environment (such as a star cluster) with a sufficiently large escape speed to retain the merger remnants~\citep{Merritt04, Gerosa:2017kvu}. Nuclear star clusters (NSCs) and gaseous active galactic nuclei (AGN) disks are the most promising sites for repeated mergers due to their larger escape speeds~\citep{Antonini:2016gqe,Antonini:2019ulv,Mapelli:2020xeq,McKernan2012MNRAS,Bartos:2016dgn,McKernan2018ApJ}. However, a subclass of globular clusters (GCs) may also facilitate hierarchical mergers of BBHs~\citep{Mahapatra:2021hme, Antonini:2022vib}. Additionally, repeated mergers offer a natural explanation for the existence of intermediate mass black holes of a few hundred solar masses~\citep{Miller:2001ez}.

The state-of-the-art $N$-body simulation codes, such as the $\tt{NBODY6}$ and $\tt{NBODY7b}$ \citep{Aarseth:2003, Wang2015}, are used to model the long-term evolution of star clusters and make theoretical predictions for the dynamical formation channel. These $N$-body simulation codes incorporate the extensive stellar evolution prescription \citep{Hurley:2000pk, Hurley2002MNRAS}, as well as the gravitational dynamics, to higher post-Newtonian accuracy \citep{Aarseth:2002ie,Aarseth:2007wv,Nitadori2012MNRAS,Aarseth:2012MNRAS}. The comparatively more-rapid
and less detailed Monte-Carlo codes such as $\tt{MOCCA}$ \citep{Giersz2013} and $\tt{CMC}$ \citep{Rodriguez2022ApJS} have been also developed to study the formation of BBHs in dense star cluster. However, neither direct $N$-body nor Monte-Carlo codes have the ability to accurately model nuclear clusters, which may get as massive as $10^8$\,M$_\odot$. 
This further motivates the use of semi-analytical models \citep{Antonini:2019ulv, Mapelli:2020xeq, KritosFast:2022arXiv}, which, while losing the intricate details and micro-physics of $N$-body models, reproduce the overall cluster evolution along with its BH population as predicted from $N$-body models.
These models allow us to venture into the realm of massive globular and nuclear star clusters.

Previous semi-analytical models studied the observed BH remnant retention probability using numerical relativity (NR) fitting formulas~\citep{Mahapatra:2021hme,Doctor:2021qfn}. They also modeled the properties of second or higher generation BHs (e.g., the mass ratio, chirp mass, spin magnitudes, effective spin parameter, etc.~\citep{Gerosa:2017kvu, Gerosa:2019zmo, Gerosa:2021hsc, Fishbach:2017dwv, Doctor:2019ruh}). These studies found that the characteristics of different BH merger generations are largely governed by the relativistic orbital and merger dynamics rather than the astrophysical environments in which they merge (see \citealt{Gerosa:2021mno} for a review of this topic). However, models of hierarchical mergers  must consider the astrophysical environment's efficiency at retaining post-merger remnants, to ensure they are available for next generation mergers.

\begin{figure}
\centering
\includegraphics[width=\columnwidth] {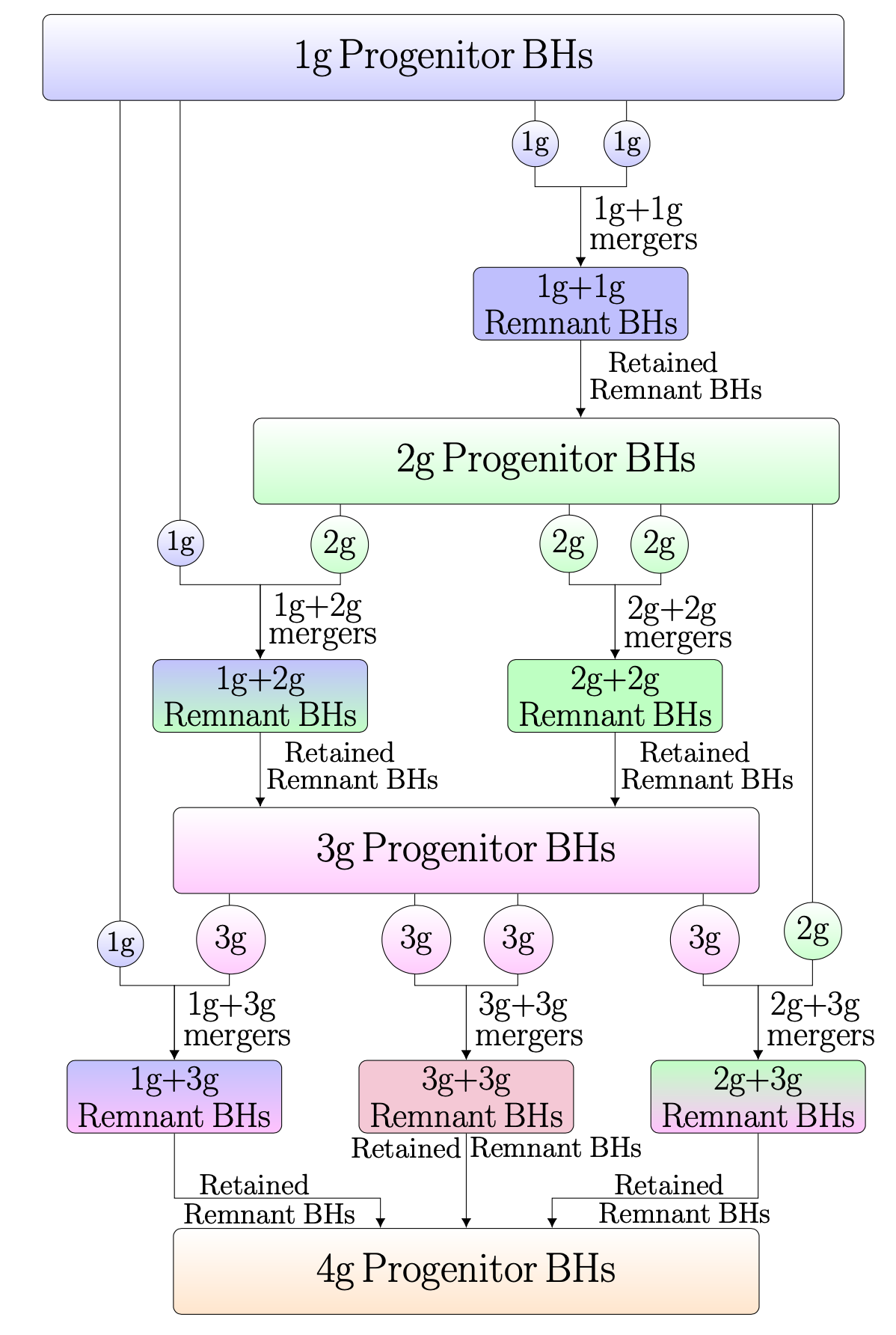}
    \caption{
     A schematic depiction of the hierarchical merger process. First generation progenitor BHs form from stellar collapse. A fraction of that seed population form the first generation BBHs (1g+1g). If their merger remnants are retained by their environment (i.e., not ejected via GW recoil), they enter the 2g progenitor BH population. A fraction of the 2g population then forms binaries with other 2g BH remnants or with members of the 1g population. The subsequent merger remnants that are retained form the 3g progenitor population, which can then form 1g+3g, 2g+3g, and 3g+3g binaries. The diagram shows the process up to the formation of the 4g progenitor population.
    }
    \label{fig:diff-gen}
\end{figure}

\citealt{Zevin:2022bfa} studied the impact of the host environment on the mass distributions of hierarchically assembled BHs. They generate merger trees that start with first-generation (1g) BH seeds, and grow them by merging BHs in series while estimating the remnants' properties using NR fits~\citep{Gerosa:2016sys}. They studied three different kinds of binary mergers: 1g+Ng, Ng+Ng, and Mg+Ng, with $\rm M \le N$. Here Ng refers to the BH generation. For example, a 1g+1g merger produces a 2g remnant, which can then form a binary with a new 1g BH (1g+2g), another 2g BH (2g+2g), or with a higher generation BH (e.g., 2g+3g; see Fig.~\ref{fig:diff-gen}).  \citealt{Zevin:2022bfa} found that once the escape speed of the host environment reaches $\sim 300\,\text{km/s}$, the fraction of hierarchically assembled binaries with total masses greater than $100 M_{\odot}$ exceeds the observed upper limit of the LVK mass distribution (see Fig.~4 of \citealt{Zevin:2022bfa}). They argued that hierarchical formation in such environments should be inhibited by some unknown mechanism to avoid this conflict termed as the ``cluster catastrophe". However, the study is hindered by several significant uncertainties that includes: (a) the initial properties of star clusters~\citep[e.g., Ref.][]{Chatterjee2017}, (b) the absence of considerations for binary-single \citep{Heggie2000MNRAS,Ginat2023} and binary-binary interactions (potentially ejecting BHs and hence decreasing the total number of hierarchical mergers~\citep[e.g., Refs.][]{Rodriguez2018,Zevin2019, Hamers2020}), (c) the absence of a varying initial BH mass function for the clusters that take into account the variation of cluster's metallicity \citep{Mapelli2013, Chattopadhyay:2022buz} and age \citep{Narloch2022}; and (d) the lack of inclusion of the change in the cluster dynamics due to the growth a massive BH at its center \citep{ArcaSedda2024}. These limitations result in numerous caveats for the study. Other works \citep[e.g., Refs.][]{Petrovich2017,Fragione2019MNRAS,Wang2021ApJ, ArcaSedda2024, ArcaSedda2024MNRAS1}, utilizing different methods, have not identified any indications of the proposed cluster catastrophe.

The main goal of this work is to develop a simple model (analogous to those developed in~\citet{Gerosa:2017kvu, Gerosa:2019zmo, Gerosa:2021hsc, Doctor:2019ruh, Gupta:2019nwj}) based on the different well-known physical phenomena (e.g., mass segregation, three body interactions) in bound environments like star clusters and calibrated with $N$-body simulations to investigate the hierarchical merger process in massive star clusters (which can retain a large fraction of higher generation mergers) and can be employed with GW data to constrain the properties of dynamically formed BBHs.
We will refer to this as a {\it Simple Parametric model for Hierarchical Mergers} or simply {\tt SPHM}. We also compare our results with those from {\tt clusterBHBdynamics} ({\tt cBHBd}), a rapid $N$-body code for evolving BH populations in dense star clusters, to assess the reliability of the {\tt SPHM}.
We then use the {\tt SPHM} to study the efficiency of hierarchical BH growth by calculating the retention probability of BBH remnants produced by different merger generations, accounting for the cluster escape speeds. This code allows us to infer the properties of multiple BBH merger generations, including their mass distributions. These properties can be computed as a function of different pairing probability functions, which depend on the total mass and mass ratio of the binaries~\citep{Pinsonneault:2005qc, Kouwenhoven:2008zb, Fishbach:2019bbm}, as well as the mass and spin distributions of the first generation progenitor BHs. Finally, we also compare the predicted mass spectrum from {\tt SPHM} with the GWTC-3 mass distribution \citep{GWTC-3-pop}. We find that the hierarchical merger scenario can explain the multiple peaks in the mass spectrum, as revealed by nonparametric population models \citep{Tiwari:2020otp,Edelman:2021zkw}. 

This paper is organized as follows. Section~\ref{sec:model} outlines in detail the assumptions that go into the {\tt SPHM}. The formation and forward evolution of BBHs are discussed in Sec.~\ref{sec:evoln}. Comparisons with the {\tt cBHBd} model is reported in Sec~\ref{sec:comparison-cBHBd}. {\tt SPHM} predictions for the efficiency of hierarchical mergers in star clusters are discussed in Sec.~\ref{sec:result}; BH mass spectrum predictions are discussed in Sec.~\ref{sec:Pm1}. Section \ref{sec:conclusion} presents conclusions and future directions.
%
%

\section{Model assumptions} \label{sec:model}
Given the component masses and spin vectors of the BBH components, {\tt SPHM} uses NR fitting formulas to predict the final mass, final spin, and kick imparted to the merger remnant due to the radiated GW energy, angular momentum, and linear momentum~\citep{Campanelli:2007ew, Lousto:2007, Lousto:2012su, Lousto:2013,Barausse:2012qz, Hofmann:2016yih}. To analyze a population of BBHs, these NR fits are supplemented with an initial mass function for the primary BH mass $m_1$ and a pairing probability function $p^{\rm pair}(m_{2}|m_{1}),$ where $m_2$ is the secondary mass.
The pairing probability phenomenologically determines the probability of forming a BBH with masses $m_1$ and $m_{2}$ in a particular cluster environment \citep{Kouwenhoven:2008zb, Fishbach:2019bbm}.
These ingredients allow us to set up and forward evolve the population through multiple generations of mergers.
 
The {\tt SPHM} starts with a population of BHs in a bound environment that, for simplicity, we refer to as a cluster. This might be a GC, a NSC, or an AGN disk. The cluster is described solely by its escape speed $V_{\rm esc}$ in our model. The initial ``first generation'' (1g) BH population is described completely by their initial mass and spin distributions. We then pair these 1g BHs via a pairing probability function $p^{\rm pair}(m_{2}|m_{1})$ to form a population of bound BBHs. (The details are described further below.) The mass, spin, and kick of the resulting BH merger remnants from this population are determined via NR fitting formulas.

Figure~\ref{fig:diff-gen} provides a schematic illustration of how {\tt SPHM} evolves the BH population. A cluster retains a merger remnant if its kick speed $V_{\rm kick}$ is less than the cluster's escape speed. Hence, the first iteration of {\tt SPHM} produces a subpopulation of second generation (2g) BHs (i.e., 1g+1g merger remnants) that are retained by the cluster. These 2g BHs are subsequently paired up with either 1g or other 2g BHs.
The second iteration of {\tt SPHM} produces 3g progenitor BHs (retained products of 1g+2g or 2g+2g mergers). A third iteration produces 4g progenitor BHs (via 1g+3g, 2g+3g or 3g+3g mergers), and so on, as the model is iterated further.   Our goal is to study, as a function of the merger generation, the fraction of binaries retained in clusters and the properties of the remnant BH population (especially its mass distribution). The {\tt SPHM} makes the following assumptions about the mass and spin distributions of the 1g BH population and the pairing probability function.

\subsection{Mass distribution}
The mass distribution of 1g progenitor BHs is assumed to follow a power-law distribution with a smooth tapering at the lower end of the distribution \citep{Kroupa:2000iv,Talbot:2018cva}: 
\begin{equation}\label{eq:IMF}
    p(m) \propto {m}^{-\alpha} \ S(m | { m_{\rm min}}, \delta_m) \mathcal{H} ({ m_{\rm max} - m}),
\end{equation}
where $\mathcal{H}$ is the Heaviside step function and $m_{\rm min}$ and $m_{\rm max}$ are the minimum and maximum masses of the power-law component of the 1g progenitor BH mass function. 
Instead of a step-function at the low-mass end of the BH mass spectrum, we use a smoothing function \mbox{$S(m \mid {m_{min}}, \delta_m)$} \citep{Talbot:2018cva} which smoothly rises from 0 to 1 over the interval $(m_{\rm min}, m_{\rm min}+\delta_{m})$:
\begin{equation}
\label{eq:smoothing}
 S(m \mid { m_{\rm min}}, \delta_m) = \begin{cases}
    0\,, \; \text{for} \;\; m< {m_{\rm min}} \,, \\
    \\
    \frac{1}{f(m - {m_{\rm min}}, \delta_m) + 1} \,,\; \\
    \;\; \text{for} \;\; {m_{\rm min}} \leq m < {m_{\rm min}} + \delta_m\,, \\
    \\
    1\,, \;\; \text{for} \;\; m\geq {m_{\rm min}} + \delta_m \,,
\end{cases}
\end{equation}
with
\begin{equation}
        f(m', \delta_m) ={\rm exp} \left( \frac{\delta_m}{m'} + \frac{\delta_m}{m' - \delta_m} \right) .
\end{equation}
Note that $\delta_m=0$ recovers the distribution with a sharp cutoff at $m=m_{\rm min}$. The spectral index $\alpha$, the tapering parameter $\delta_m$, and the minimum and maximum allowed masses $m_{\rm min}$ and $m_{\rm max},$ fully define the mass distribution of 1g progenitor BHs. These parameters depend on the initial stellar mass function, the stellar evolution process (e.g., wind prescription, natal kick distribution, etc.), and the cluster properties (such as metallicities, cluster mass, and virial radius).

\subsection{Spin distribution}
The uncertainties in the spin magnitudes of BHs from stellar evolution and core-collapse supernova models are still too large to predict the 1g progenitor BH spin distributions. Efficient angular momentum transport via the magnetic Tayler instability leads to low birth spins for BHs~\citep{Spruit2002, Fuller2019MNRAS, Fuller:2019sxi}. Here, we assume a uniform distribution between $[0, 0.2]$ for the dimensionless spin magnitude $\chi$ of 1g BHs. Spin tilt angles for all BH generations and population models are isotropically drawn over a sphere~\citep{Rodriguez2016, Tagawa:2020dxe}. Note that spin tilt angles are chosen at binary separations $\sim 10M,$ where NR simulations typically start (see \citealt{Barausse2009ApJ} and Sec.~2.1 of \citealt{Gerosa:2014gja}).

\subsection{Pairing function}\label{sec:pairing-functions}
Dynamical interactions in dense clusters produce more equal-mass mergers (see Sec.~IVB and Fig.~9 of \citealt{Rodriguez2016}). For binary-single and binary-binary encounters in dense clusters, binaries are prone to exchanging components, preferentially expelling less-massive components in favor of more massive companions \citep{Sigurdsson1993Nature,Heggie1975MNRAS}. Additionally the merger probability in dense clusters depends on the total mass ($M_{\rm tot}$) of the binary as dynamical interactions favor binaries with larger $M_{\rm tot}$ due to the well known effect of ``mass segregation''~\citep{Chandrasekhar1943ApJ,Spitzer1987}. Binaries merging in AGN disks typically have lower mass ratios (i.e., $q\equiv m_2/m_1 \ll 1$) as migration traps within AGN disks give rise to interactions that lead to unequal-mass binaries (\citealt{McKernan2012MNRAS, Bellovary:2015ifg}; see also Fig.~3 of \citealt{Yang:2019okq}, as well as Figs.~3 and 5 of \citealt{Li:2022gul}).

Here we consider three types of pairing probability functions: one that depends only on the mass ratio $q$, one that depends only on the total mass $M_{\rm tot}$, and one that depends on both $q$ and $M_{\rm tot}$, where, as before, $q\equiv m_2/m_1 \leq 1$. In the first case, we assume the pairing probability depends only on the mass ratio via \citep{Fishbach:2019bbm}
\begin{equation}{\label{eq:Pair-q}}
    p^{\rm pair}(m_{2}|m_{1}) \propto q^{\beta} \,,
\end{equation}
where $\beta$ ($\ge 0$) is the sole parameter that governs the pairing probability between two BHs in a cluster. For positive values of $\beta$, the pairing function in Eq.~(\ref{eq:Pair-q}) favors the formation of comparable-mass binaries. But it does not account for mass segregation. 

Next, we consider the pairing function that depends only on the total mass via \citep{O'Leary2016ApJ,Mapelli:2021syv}
\begin{equation}{\label{eq:Pair-Mtot}}
    p^{\rm pair}(m_{2}|m_{1}) \propto M_{\rm tot}^{\beta} \,.
\end{equation}
For $\beta>0$, the pairing function in Eq.~(\ref{eq:Pair-Mtot}) prefers the formation of massive binaries and hence captures the mass segregation.

Finally, we also consider the pairing probability function that depends on both $q$ and $M_{\rm tot}$:
\begin{equation}{\label{eq:Pair-q-Mtot}}
    p^{\rm pair}(m_{2}|m_{1}) \propto q^{\beta_1} M_{\rm tot}^{\beta_2} \,.
\end{equation}
This pairing function captures the preference for forming both equal-mass binaries and massive BH binaries in the cores of dense clusters for $\beta_{1,2}>0$. In all of the three cases above [Eqs.~(\ref{eq:Pair-q})\mbox{--}(\ref{eq:Pair-q-Mtot})] the normalization factor depends on $m_{\rm min}$ and $m_{\rm max}$ via the condition $\int_{m_{\rm min}}^{m_{\rm max}} dm_1 \int_{q_{\rm min}}^{1} dq \, p^{\rm pair} =1$, where $q_{\rm min}=\tfrac{m_{\rm min}}{m_{\rm max}}$.
For all the pairing functions, $\beta=0$ ($\beta_1=\beta_2=0$ for Eq.~(\ref{eq:Pair-q-Mtot})) corresponds to random pairing \citep{Kouwenhoven:2008zb} and preferences the formation of unequal-mass binaries.

The pairing of BHs in dense clusters depends on the cluster properties (e.g., metallicities, evolution history, initial  cluster mass, and virial radius) and different types of dynamical encounters inside the cluster core~\citep{Antonini:2022vib}. The parameter $\beta$ captures these physical processes and is (in principle) a function of the above-mentioned cluster parameters. Here we assume that the cluster properties (along with $V_{\rm esc}$) do not evolve over the timescale of the entire hierarchical merger process. In reality, the pairing probability may depend on the cluster's dynamical age \citep{Chattopadhyay:2022buz, Antonini:2019ulv, Antonini:2022vib}. By the time much higher generation BBHs have formed (e.g., 10th generation), the cluster should have evolved significantly from its initial stage. In that case, one should incorporate the evolution of the star clusters in the computation. For example, after 1 Gyr of evolution, the cluster compactness (compactness is defined as the ratio of the total cluster mass and its half-mass radius) reduces by a factor of $\sim 5$ (see Fig.~(7) of \cite{Antonini:2019ulv}). The escape speed of a cluster is proportional to the square root of its compactness (See Eq.~(29) of \cite{Antonini:2019ulv}). Therefore, after 1 Gyr of evolution, the escape speed of a cluster reduces by $\sim 45\%$. On the other hand, the hierarchical growth of BHs in massive clusters such as NSCs can reach up to as high as the 8th generation (See Fig.~(12) of \cite{Mapelli:2021syv}) within 1 Gyr of evolution. Here, we restrict ourselves to the formation of third-generation BBHs.

\subsection{Other assumptions}
The {\tt SPHM} also assumes that mutli-body interactions play a negligible role in ejecting BHs and altering the fraction of BHs that are retained.  Dynamical ejection is an important effect in clusters with low escape speeds such as young star clusters  
\citep{Banerjee2010mnras} and GCs. For example, simulations of Milky Way type GCs ($V_{\rm esc} \sim 30$ km/s) found that roughly $\sim50\%$ of cluster BBHs merge \emph{after} being ejected from the cluster by dynamical 3-body interactions (see Sec.~9.1 of \citealt{Kremer:2019iul}). 
However, GCs are more massive at their birth \citep{Webb2015MNRAS} and would have retained most of their BBHs. Additionally, \citet{Miller:2008yw} and \citet{Antonini:2016gqe} argued that dynamical ejections in NSCs can be safely ignored due to their larger escape speeds (see also \citealt{Antonini:2018auk}; at present day the GCs have escape speeds $\sim2-180$ km/s, whereas the NSCs have escape speeds $\sim10-1000$ km/s). For astrophysical environments that facilitate hierarchical mergers (those with $V_{\rm esc}\ge200$ km/s; \citealt{Mahapatra:2021hme}), ignoring dynamical ejections is a reasonable assumption. The position of the BBH within the cluster, along with any sinking due to dynamical friction, are likewise ignored. Inclusion of these effects will reduce the efficiency of hierarchical mergers; hence our estimates should be seen as upper limits.

\section{Evolving the BBH population}\label{sec:evoln}
To evolve our BBH population we begin by randomly drawing $\sim 10^6$ BH pairs from the previously described mass and spin distributions. Each pair is characterized by mass and spin parameters \{$m_1^{\rm 1g}$, $m_2^{\rm 1g}$, $\boldsymbol{\chi}_{1}^{\rm 1g}$, $\boldsymbol{\chi}_{2}^{\rm 1g}$\}, with subscript ``1'' denoting the heavier ``primary.''
A pool of 1g+1g binaries inside a cluster with escape speed $V_{\rm esc}$ is then constructed by sampling over these pairs with a weight proportional to $p^{\rm pair}$ [using either Eq.~(\ref{eq:Pair-q}), (\ref{eq:Pair-Mtot}), or (\ref{eq:Pair-q-Mtot})].

We then estimate the final mass $M_{\rm f}$, spin $\chi_{\rm f}$, and kick $V_{\rm kick}$ of each binary's merger remnant using NR fits. (These fits were developed in \citealt{Barausse:2012qz,Hofmann:2016yih,Campanelli:2007ew} and are summarized in Sec.~V of \citealt{Gerosa:2016sys}; see also Appendix A of \citealt{Mahapatra:2021hme}). This produces the first generation 1g+1g merger remnants. If a 1g+1g remnant is retained in the cluster, it can take part in further mergers. The necessary condition for the retention of a remnant in a cluster is that the GW kick imparted to the remnant should be less than the escape speed of the host cluster ($V_{\rm kick}<V_{\rm esc}$). Hence, the probability of repeated mergers in a cluster is directly proportional to the retention probability: the probability that a member of the population has $V_{\rm kick}<V_{\rm esc}$.
\citet{Mahapatra:2021hme} estimated the retention probability of GWTC-2 BBH events inside clusters with different escape speeds via the cumulative distribution function (CDF) of the kick posteriors (see Sec.~2 of \citet{Mahapatra:2021hme}). Here the retention probability of 1g+1g mergers is calculated from the kick CDF of the population of 1g+1g merger remnants. The retention probability for 1g+1g remnants (and for future merger generations) can then be quantified as a function of the cluster escape speed.
The retained 1g+1g remnants will produce the population for the second generation (2g) progenitor BHs. 

There are two possibilities for second generation mergers: 1g+2g or 2g+2g binaries that are formed according to the pairing probability function. The properties and retention probabilities of the resulting mergers are computed as described above, with further details in the Sec.~\ref{sec:step-appen} of Supplemental Material.
The retained remnants form the third generation (3g) progenitor BH population. From there, the third generation of binaries is paired via three possible combinations: 1g+3g, 2g+3g, and 3g+3g. The subpopulation  that is retained after these 3g binaries merge forms the 4g progenitor BH population. The Sec.~\ref{sec:step-appen} of Supplemental Material describes the steps of the forward evolution in an algorithmic way. In principle, this process goes on until there are no more BHs left in the cluster and/or inefficiency in pairing halts the process. Considering the limited number of high-mass BHs in the current LVK observational sample, we stop our iteration when 4g progenitor BHs are formed. Our forward-evolution model of the BBH population closely follows \cite{Gerosa:2017kvu, Gerosa:2019zmo, Gerosa:2021hsc, Doctor:2019ruh, Gupta:2019nwj}.

\begin{figure*}
    \includegraphics[width=1.0\textwidth]{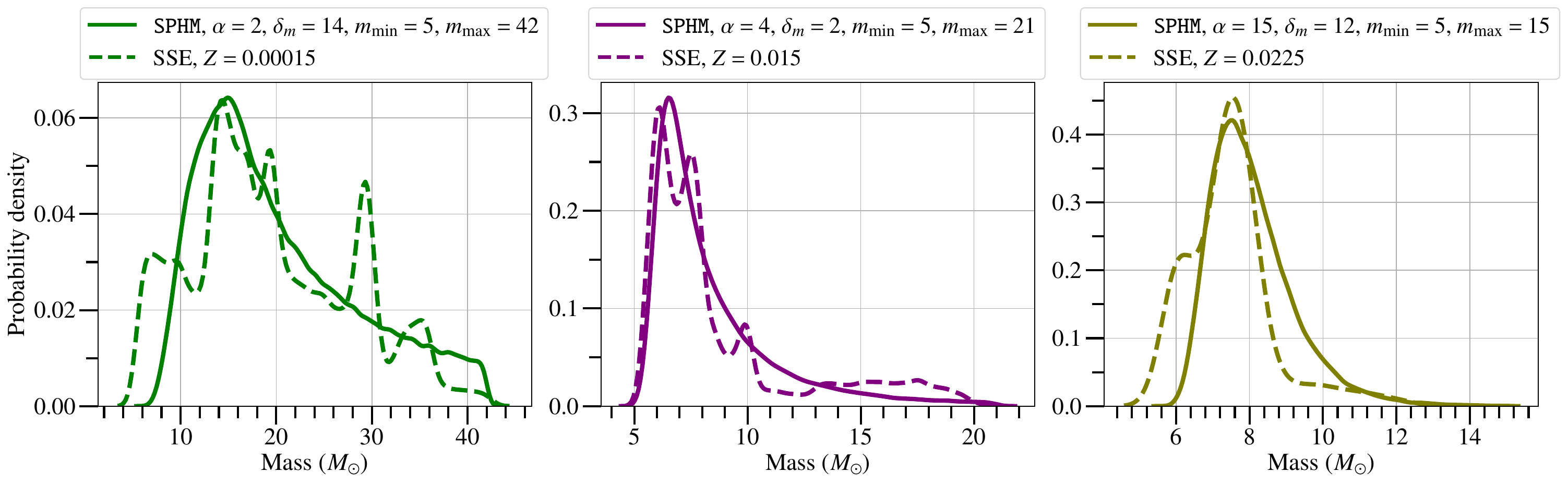}
    \caption{
    The initial mass distribution of BHs obtained from the SSE code (dashed curves), used in {\tt cBHBd}, and the fitted initial mass distributions to the SSE data obtained by varying the free parameters in Eq.~(\ref{eq:IMF}) (solid curves), used in {\tt SPHM},  are shown.
    We consider three values for cluster metallicity: $Z=0.00015$ (left panel), $Z=0.015$ (middle panel), and  $Z=0.0225$ (right panel). For each value of cluster metallicity, the fitted parameters of Eq.~(\ref{eq:IMF}) to the SSE data are shown in the legend where $\delta_m$, $m_{\rm min}$, and $m_{\rm max}$ are in $M_\odot$. 
   }
    \label{fig:IMF-fit}
\end{figure*}

\section{Comparison with cBHBd}\label{sec:comparison-cBHBd}
In this section, we compare the prediction of the mass spectrum for various BBH generations from the {\tt SPHM} with the fast semi-analytic cluster population model {\tt cBHBd}~\citep{Antonini:2019ulv,Antonini:2022vib}.

\subsection{cBHBd model}\label{sec:cbhbd} 
The semi-analytical code $\tt{cBHBd}$ \citep{Antonini:2019ulv,AntoniniGieles2020,Antonini:2022vib,Chattopadhyay:2023arXiv} relies on Henon's principle~\citep{Henon:1972}, which states that---following core-collapse---the rate of heat generation in the cluster's core is a constant fraction of the total cluster energy per half-mass relaxation time. In the ``balanced evolution'' phase of the cluster, the heat is produced by the hard BH binaries in the core~\citep{Breen_Heggie:2013}.
The evolution of the BH population---which primarily segregates to the cluster core---is therefore linked with the global properties of the host cluster, since the hardening of BH binaries through dynamical interactions in the core acts as a heat source for the cluster.

The $\tt{cBHBd}$ model approximates the cluster as a two-component system i.e., the cluster is composed of two types of members: (a) BHs and (b) all other stellar remnants and remaining small still-evolving stars~\footnote{This is a reasonable assumption because the time scale of stellar evolution $\sim \mathcal{O}\mathrm{(Myrs)}$ is much smaller compared to the time scale for cluster evolution $\sim\mathcal{O}\mathrm{(100\, Myrs)}-\mathcal{O}\mathrm{(Gyrs)}$~\citep{Hurley:2000pk}.
}. It further assumes that the cluster's primary heat source is a single BH-BH binary at the core hardened through binary-single encounters. It is also assumed that the binding energy of the binary increases by a fraction of 0.2 during the binary-single interactions.
After each encounter of the double BH with a single BH, the fractional decrease of the binary semi-major axis is computed and checked if the double BH merges inside or outside the cluster after getting ejected out.
If the BH binary merges inside the cluster, its recoil kick is also calculated to check if the remnant is retained to further partake in forming BH pairs. 
Simultaneously, the cluster's global properties are also evolved from its initial state. The initial conditions of the cluster are specified by three parameters: the cluster mass, the cluster density, and the total mass of all BHs in the cluster. The latter is derived using the realistic initial mass function of stars and the single stellar evolution code. The detailed steps of the time evolution of the cluster properties are provided in Sec.~(2.3) of \citealt{Antonini:2019ulv}. The generation of the initial BH mass function, the double BH pairing prescription and BH binary-single encounters are chalked in the paragraphs below. Therefore, adopting a reliable model for cluster evolution, realistic initial BH mass distribution, and model of BH dynamics in clusters, $\tt{cBHBd}$ predicts double BH formation, evolution, and merger rates in star clusters.

The masses of the BH stellar progenitors are sampled from the Kroupa initial mass function, $p(m_{\star}) \propto m_{\star}^{-2.3}$~\citep{Kroupa:2000iv,Salpeter1955ApJ}, with $m_{\star}$ corresponding to initial stellar masses in the range $20 M_{\odot}\mbox{--}130 M_{\odot}$. For a given cluster metallicity $Z$, individual stars are evolved into BHs using the Single Stellar Evolution (SSE) code~\cite{Hurley:2000pk}. The original version of SSE is publicly available with updates at \url{https://astronomy.swin.edu.au/~jhurley/}~\citep{SSE2013}. In this study, we employ a modified version of the SSE code incorporating several key updates. The masses of compact objects are assigned following the prescription of~\citet{Belczynski2008}. The mass loss treatment from stellar winds adopts the updated model of~\citet{Vink2001}, as implemented in~\citet{Belczynski2010}. The mapping of pre-supernova progenitor core mass to BH mass and the scaling of BH natal kick magnitudes (the BH natal kick is drawn from the Maxwellian distribution with a dispersion of 265~km/s~\citealt{Hobbs2005MNRAS}) with fallback mass~\cite{Janka:2013hfa}, follows the framework of~\citet{Fryer2012ApJ}. The ejection of BHs due to natal kicks is also accounted for following~\citet{Hobbs2005MNRAS}. The modifications to the original SSE code described above have been used in~\citet{Chattopadhyay:2022buz}. Furthermore, the (pulsational) pair-instability mass loss and upper mass gap~\citep{Woosley2017ApJ} prescriptions from~\citet{Spera:2017fyx} are incorporated to estimate the initial mass distribution of BHs. This enhanced version of the SSE code, which has also been used in recent works such as~\citet{Antonini:2022vib, Chattopadhyay:2023arXiv}, is employed here.

The pairing probabilities for BH binaries and binary-single encounters are given by a metallicity-dependent initial BH mass function that is a power law distribution; see \citet{Antonini:2022vib} and \citet{Chattopadhyay:2023arXiv} for more details.\footnote{If two BHs with masses $m_1$ and $m_2$ form a binary, the probabilistic pairings are given by $p(m_1) \propto m_1^{\lambda_1}$ (which determines which massive BH $m_1$ is likely to be in a binary) and $p(q) \propto q^{\lambda_2}$ (which dictates the mass-ratio of the binary BH with $m_1$ as primary, hence picking the value of $m_2$) with $\lambda_1=8 + 2\lambda$ and $\lambda_2=3.5 + \lambda$. 
The probability of this binary encountering another BH with mass $m_3$ is given by $p(m_3) \propto m_3^{\lambda_3}$, with $\lambda_3=0.5 + \lambda$. Here, $\lambda$ is an eighth-order polynomial fit to the cluster metallicity as given by \citet{Chattopadhyay:2023arXiv}.
} The binary energy loss (via hardening from binary-single encounters) depends on the single-to-binary mass ratio \citep{Hills_Fullerton:1980, Quinlan:1996, Chattopadhyay:2023arXiv}. The $\tt{cBHBd}$ model accounts for the dynamical ejection of BHs through binary-single encounters, as well as binary hardening and merger through binary-single encounters as detailed in \citet{Antonini:2022vib} and \citet{Chattopadhyay:2023arXiv}.
While binary-binary encounters are not accounted for, \citet{Barber:2023hvo} has shown that clusters with initial escape speeds $\gtrsim100$\,km/s have negligible binary-binary interactions and are dominated by binary-single encounters. Moreover, the simultaneous evolution of the host cluster and its BH population in ${\tt cBHBd}$ has been assessed against direct $N$-body models \citep{Antonini:2019ulv}. 

\subsection{Comparison setting}\label{sec:comparison-setting}
Here, we have adopted data from the {\tt cBHBd} model for two different values of the cluster metallicity, $Z=0.00015$, and $Z=0.0225$. The initial cluster mass is $M_{\rm cl,i}=2 \times 10^{7}\, M_{\odot}$ and the initial half-mass density is $\rho_{\rm h,i}=10^{7}\,M_{\odot} \, {\rm pc}^{-3}$.
This chosen model is the ``fiducial'' of Ref.~\cite{Chattopadhyay:2023arXiv}, which roughly matches the Milky Way nuclear cluster after one Hubble time of evolution \citep[see Fig. 19 of Ref.][]{Chattopadhyay:2023arXiv}\footnote{Since we assume the model to be a nuclear cluster, only stellar evolution mass loss and cluster expansion due to core dynamics is included; mass-loss due to stripping by a host galaxy is not included.}. 
The cluster mass $M_{\rm cl,0}$ post mass-segregation (at core collapse) is calculated to be about half of the cluster's initial mass, $M_{\rm cl,0}\approx 10^{7}\, M_{\odot}$, while the half-mass density at core-collapse is found to be $\rho_{\rm h,0} \approx 2 \times 10^{6}\,M_{\odot} \, {\rm pc}^{-3}$.
Therefore, the net escape speed of the cluster at the onset of the dynamical regime (post-core-collapse) is $\sim 400$ km/s (calculated using Eq.~(29) of \citet{Antonini:2019ulv}). For comparison with the {\tt SPHM}, we have extracted the mass distributions of 1g+1g, 1g+2g, and 1g+3g BBH mergers from the {\tt cBHBd} model for the two different values of the cluster metallicity mentioned above.
We likewise generate the same mass distributions using the {\tt SPHM}. We first calculate the initial mass distribution of BHs inside a cluster with an escape speed of 400 km/s corresponding to the {\tt cBHBd} data. The cluster members (i.e., stellar and BH populations inside the cluster) are obtained from SSE assuming a Kroupa initial mass function and the previously mentioned stellar evolution processes.
While estimating the initial mass distribution of BHs, we incorporate the ejection of BHs from the cluster by natal kicks. Once we get the data for the initial mass distribution of BHs inside the cluster, we fit this to the function in Eq.~(\ref{eq:IMF}).
The fitted initial mass distribution and the actual initial mass distribution obtained from the SSE code are shown in Fig.~\ref{fig:IMF-fit}. We find that Eq.~(\ref{eq:IMF}) with parameters $\alpha=2,  \, \delta_{m}=14M_\odot, \, m_{\rm min}=5M_\odot, \,{\rm and} \, m_{\rm max}=42M_\odot$ provides a reasonably good fit to the initial mass distribution of BHs in a cluster with metallicity $Z=0.00015$. Similarly, we also obtain the set of parameters that gives a good fit for high metallicity initial mass functions for $Z=0.015, 0.0225$. The best-fit parameters are presented in Fig.~\ref{fig:IMF-fit}. 

Next, we adopt these values of $\alpha, \, m_{\rm min}, \, m_{\rm max}, \,{\rm and} \, \delta_{m}$ (with $V_{\rm esc}=400$ km/s) in {\tt SPHM} and produce the mass distribution of different BBH generations for the three different types of pairing probability functions (discussed in Sec.~\ref{sec:pairing-functions}).
For each pairing probability function, we consider a list of values for the pairing exponents ($\beta$, or, $\beta_1$ and $\beta_2$) by varying their value from 0 to 15 in steps of 0.1. Note that for the pairing function of Eq.~(\ref{eq:Pair-q-Mtot}), a two-dimensional grid ranging from 0 to 15 in both directions with a uniform step size of 0.1 is constructed for ($\beta_1,\, \beta_2$). We then compute the mass spectrum for each value of $\beta$ or ($\beta_1,\, \beta_2$) and compare the distributions of primary mass and mass ratio from the {\tt SPHM} with the {\tt cBHBd} model for 1g+1g, 1g+2g, and 1g+3g BBH mergers for the two types of cluster metallicities.

For each considered value of the pairing exponents, we further estimate the Jensen–Shannon (JS) divergence~\cite{JSdiv} for the mass spectrum of 1g+1g, 1g+2g, and 1g+3g BBH mergers between the SPHM and cBHBd models. The JS divergence is a useful measure to quantify differences between probability distributions. The smaller the value of the JS divergence, the greater the agreement between two probability distributions. For each pairing function, we note down the values of the pairing exponents for which the JS divergences between the SPHM and cBHBd models are the smallest. The corresponding values represent the best-fit values for the pairing exponents within the accuracy of the employed step size of 0.1. We report our findings from these comparisons in the next section.

\begin{figure*}[ht]
    \includegraphics[width=0.95\textwidth]{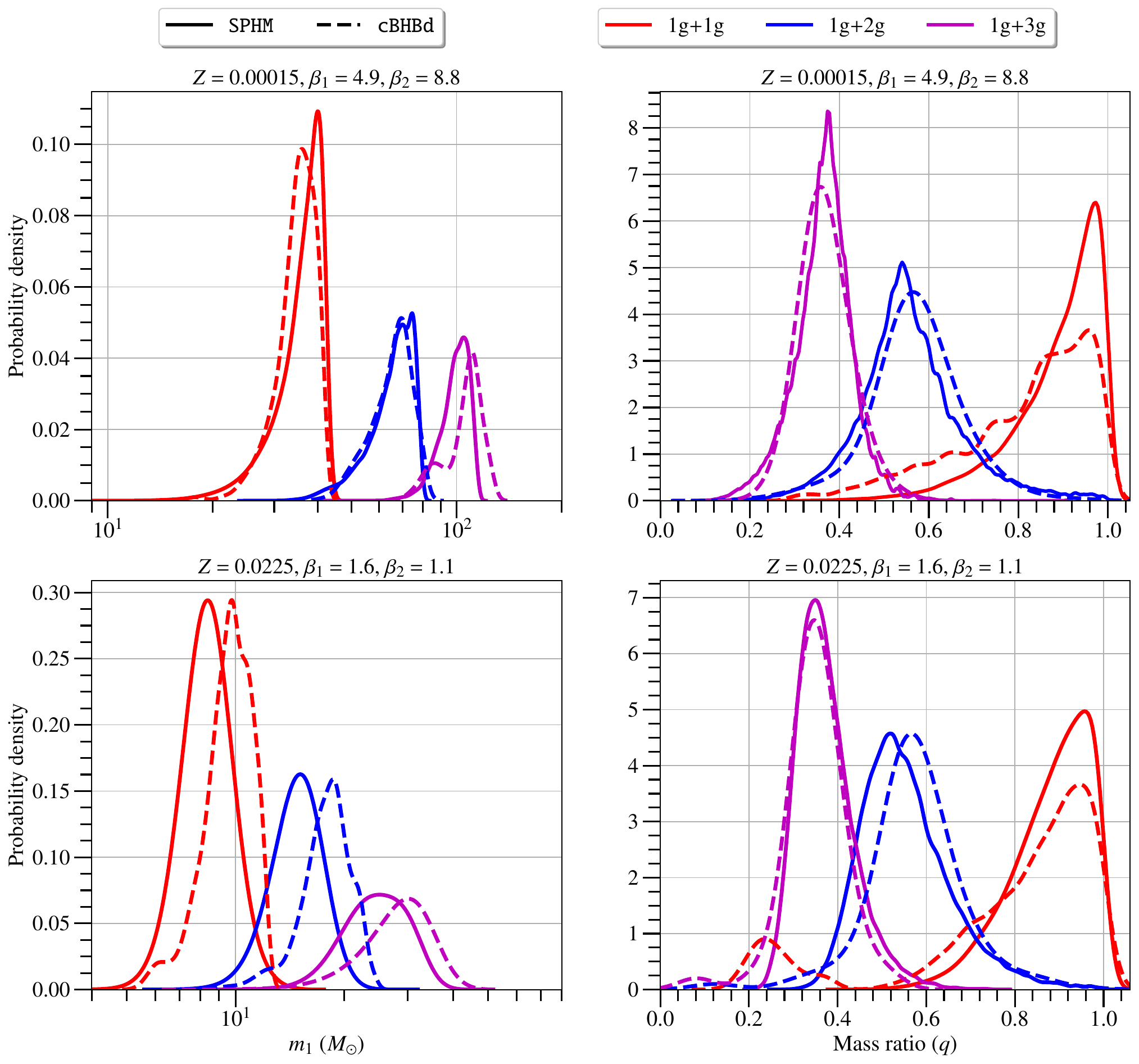}
    \caption{Comparison of the BBH distributions for the binary's primary mass (first column) and mass ratio (last column) for different BBH generations as computed via the {\tt SPHM} and {\tt cBHBd} models. Here, the {\tt SPHM} assumes the $M_{\rm tot}-q$-based pairing function in Eq.~(\ref{eq:Pair-q-Mtot}). The different colored curves show the different generations of BBH mergers. The solid-colored curves are derived from the {\tt SPHM} [with $(\beta_1,\, \beta_2)$-values as  indicated in the plot titles], while the dashed-colored curves are derived from the {\tt cBHBd} model (for the indicated metallicity values).
    These choices of $(\beta_1,\, \beta_2)$ are the best-fit values with an accuracy of 0.1. See the text in Sec.~\ref{sec:comparison-setting} and Sec.~\ref{sec:comparison-q-Mtot-pair} for more details. We see that the results of the {\tt SPHM} are in reasonably good agreement with {\tt cBHBd} results overall.
   }
    \label{fig:m1-q-comparison--SPHM-cBHBd}
\end{figure*}

\begin{figure*}[ht]
    \includegraphics[width=0.95\textwidth]{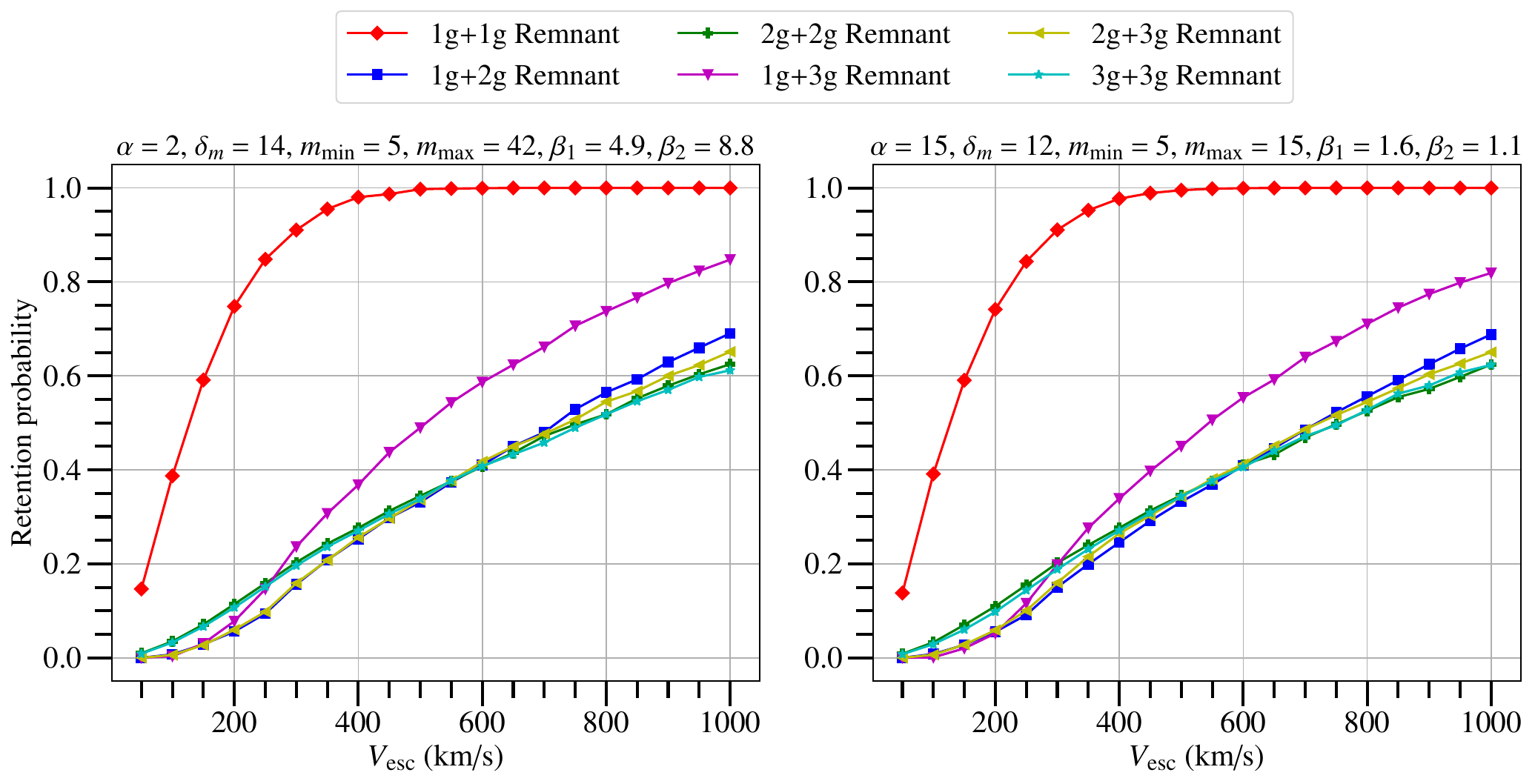}
    \caption{Retention probabilities as a function of cluster escape speed $V_{\rm esc}$ for different kinds of merger remnants. The different curves in each panel indicate the various possible binary merger combinations (i.e., 1g+2g, 2g+2g, etc.; see top legend.) Each panel uses different population model assumptions for the 1g progenitors. These plots assume the pairing function in Eq.~(\ref{eq:Pair-q-Mtot}). The retention probability of a particular merger remnant type is computed directly from the kick CDF of that population as discussed in the text and the Sec.~\ref{sec:step-appen} of Supplemental Material.
    Retention probability decreases as we go from first generation mergers to higher generation mergers.
    However, note that the retention probability of 1g+2g remnants is always larger than 1g+3g remnants for $V_{\rm esc}<200$ km/s whereas the retention probability of 1g+3g remnants is always larger than 1g+2g remnants for $V_{\rm esc}\gtrsim200$ km/s.
    For mergers that involve a particular generation, unequal mass binaries have higher retention probabilities than equal mass binaries (e.g., $\rm{P_{13}^{ret}}>\rm{P_{23}^{ret}}>\rm{P_{33}^{ret}}$, where $\rm P^{\rm ret}_{\rm MN}$ refers to the retention probability of the remnant of a Mg+Ng binary merger) in clusters with $V_{\rm esc}\gtrsim500$ km/s. The values of $\delta_m$, $m_{\rm min}$, and $m_{\rm max}$ in the titles are in $M_\odot$. 
}
    \label{fig:Ret-prob}
\end{figure*}

\subsection{Results of the comparison for different pairing functions}\label{sec:comparison}
We now compare the results of the mass distributions of different generations of mergers from {\tt SPHM} with {\tt cBHBd} for different pairing functions. Though we have performed extensive comparisons, for the convenience of presentation, we show only those results that give good agreement between the two models. A brief discussion of the level of disagreements and potential reasons are summarized for each pairing.

\subsubsection{Pairing based on mass ratio}
We first compare the primary mass and the mass ratio distributions for different BBH generations as computed via the {\tt SPHM} and {\tt cBHBd} models, where {\tt SPHM} assumes the $q$-based pairing given in Eq.~(\ref{eq:Pair-q}).
For the low metallicity-case ($Z=0.00015$), the mass spectrum of different BBH generations from the {\tt SPHM} does not agree with the results of {\tt cBHBd}; there is a marginal agreement for $Z=0.0225$. In low metallicity clusters, the initial BH mass distribution is broad (between $\approx 5 M_{\odot}\mbox{--}45 M_{\odot}$ for $Z=0.00015$); there the effect of mass segregation allows only the massive BHs to sink into the cluster core and form binaries. As the $q$-based pairing does not account for the mass segregation, the {\tt SPHM} with $q$-based pairing does not agree with the {\tt cBHBd} model when $Z=0.00015.$ In high metallicity clusters, the initial mass functions of BHs are relatively narrower (between $\approx 5 M_{\odot}\mbox{--}15 M_{\odot}$ for $Z=0.0225$). In that case, the mass segregation is less significant as there is not much diversity in the range of initial BH masses. Therefore, the {\tt SPHM} with $q$-based pairing marginally agrees with the {\tt cBHBd} model when $Z=0.0225$. Here, $\beta=1.6$ provides the smallest JS-divergence.
\subsubsection{Pairing based on total mass}
Next, we make the same comparisons between the two models when the {\tt SPHM} assumes the $M_{\rm tot}$-based pairing function in Eq.~(\ref{eq:Pair-Mtot}).
Here, we find that the primary mass distributions from the {\tt SPHM} agree well with the {\tt cBHBd} model for metallicity $Z=0.00015$ when $\beta=9.9$ (provides the smallest JS divergence).
However, the mass ratio distributions from these two models do not agree with each other for $Z=0.00015$. For $Z=0.0225$, there is again marginal agreement between the two models. In this case, $\beta=0.6$ yields the smallest value for the JS divergence.
Moreover, for high metallicity clusters, the type of pairing does not yield significantly different results. This is unsurprising, given that these clusters produce relatively lower-mass BHs, resulting in a narrow initial BH mass spectrum. As a result, both models tend to agree with each other.

\subsubsection{Pairing based on mass ratio and total mass}\label{sec:comparison-q-Mtot-pair}
Lastly, we compare the mass spectrum of {\tt SPHM} and {\tt cBHBd} model when the {\tt SPHM} uses the pairing probability function given by Eq.~(\ref{eq:Pair-q-Mtot}). Here, we find good agreement between these two models for both the metallicities.
This is because the pairing function in Eq.~(\ref{eq:Pair-q-Mtot}) captures both the formation of equal-mass binaries from three-body interactions and the mass segregation effect to a greater extent than the previous two pairing functions.
Here, ($\beta_1=4.9$, $\beta_2=8.8$) and ($\beta_1=1.6$, $\beta_2=1.1$) provide the best agreement between the {\tt SPHM} and {\tt cBHBd} models for $Z=0.00015$ and $Z=0.0225$, respectively.
\footnote{As our goal in this section is to assess the agreement between the {\tt SPHM} and the {\tt cBHBd} model, we considered only some representative values of the model exponents and have not attempted to see how these exponents map onto microphysical parameters such as metallicity.} We have shown the agreement between these two models in Fig.~\ref{fig:m1-q-comparison--SPHM-cBHBd}.

Independently, we have also forward evolved the data of the initial mass distribution of BHs inside the cluster (directly derived from the SSE package with various stellar physics inputs mentioned in the first paragraph of Sec.~\ref{sec:comparison-setting}) using the {\tt SPHM} with the three pairing functions. For the pairing functions in Eq.~(\ref{eq:Pair-q-Mtot}), we also find good agreement between the {\tt SPHM} and the {\tt cBHBd} model for both the metallicities, $Z=0.00015$ and $Z=0.0225$.

To summarize, we find that good agreements in the mass spectrum and the mass ratio distributions from the {\tt SPHM} with the pairing function of Eq.~(\ref{eq:Pair-q-Mtot}) when compared against the {\tt cBHBd} code for representative cases of low and high metallicities. Having established this agreement, we will use only the {\tt SPHM} model with the pairing function of Eq.~(\ref{eq:Pair-q-Mtot}) from now on.

\section{Efficiency of hierarchical mergers} \label{sec:result}
 In this section we study the retention probability for BBH mergers as a function of a cluster's escape speed and for two sets of model parameters.
 We choose $\alpha=2$, $\delta_{m}=14M_\odot$, $m_{\rm min}=5M_\odot$, and $m_{\rm max}=42M_\odot$ for the 1g BH progenitor mass function and ($\beta_1=4.9,\, \beta_2=8.8$) for the pairing exponents in Eq.~(\ref{eq:Pair-q-Mtot}) in the first set of model parameters. The first set is the representative of metal-poor clusters with $Z\sim0.00015$. In the second set, we choose $\alpha=15$, $\delta_{m}=12M_\odot$, $m_{\rm min}=5M_\odot$, and $m_{\rm max}=15M_\odot$ for the 1g BH progenitor mass function and ($\beta_1=1.6,\, \beta_2=1.1$) for the pairing exponents in Eq.~(\ref{eq:Pair-q-Mtot}). The second set of model parameters is the representative of metal-rich clusters with $Z\sim0.0225$.
 Using these model parameters, we forward evolve the BBH populations with the {\tt SPHM} assuming the pairing function in Eq.~(\ref{eq:Pair-q-Mtot}). We estimate the retention probability of different merger generations in clusters as a function of the cluster escape speed $V_{\rm esc}$. The calculation of the retention probability is explained in the second paragraph of Sec.~\ref{sec:evoln}.

Figure \ref{fig:Ret-prob} shows the retention probability versus cluster escape speed for different types of mergers. The figure suggests that a cluster with a $400$ km/s escape speed has a retention probability between 25\% and 95\%,  depending on the merger type and the assumptions about the 1g progenitor population. Among mergers that involve a particular generation (e.g., 2g+2g and 1g+2g, which both have 2g BHs as components), unequal mass mergers have higher retention probabilities than equal mass mergers in clusters with $V_{\rm esc}\gtrsim500$ km/s. For example, in second generation mergers 1g+2g mergers have higher retention probabilities than 2g+2g mergers in clusters with $V_{\rm esc}\gtrsim500$ km/s. This is due to the lower kick speeds in asymmetric BBH mergers. We find that the retention probabilities of merger remnants fall off as we proceed from first-generation mergers to higher-generation mergers due to larger spin magnitudes of the higher-generation BHs. However, the retention probabilities increase for asymmetric mergers in higher generations. For example, 1g+3g mergers have greater retention probabilities than 1g+2g mergers in most of the cases.

A crucial question is how abundant are star clusters that have escape speeds of at least $400$ km/s. From current observations the escape speed distributions at present-day ($z=0$) for GCs and NSCs peak at $\sim 30$ km/s and $\sim 150$ km/s, respectively \citep{Antonini:2016gqe, Georgiev:2016, Harris:1996}. (See Fig.~3 of \citealt{Antonini:2016gqe}; note that these escape speed estimates are from the cluster center for GCs, whereas for NSCs they are defined at the half-mass cluster radius.) Using those present-day escape speed distributions, we find that the retention probability of merger remnants falls off from $\sim 61\%$ $(\sim 10\%)$ to $\sim 13\%$ ($\sim 1\%$) as we proceed from first generation mergers to higher generation mergers in NSCs (GCs). However, star clusters were more massive and denser at higher redshifts; the corresponding higher escape speeds will increase the retention probability. For instance, at birth GCs have masses that are on average a factor of $\sim 4.5$ times larger than their present day masses \citep{Webb2015MNRAS}, increasing their escape speeds by a factor $\sim \sqrt{4.5}\approx 2.1$. Our current understanding of the redshift evolution of clusters is limited.
Future observations may help us map the hierarchical merger efficiency vs.~redshift if we can confidently identify a subpopulation of mergers as having hierarchical origins.


\begin{figure}[ht]
   \centering
    \includegraphics[scale=0.44]{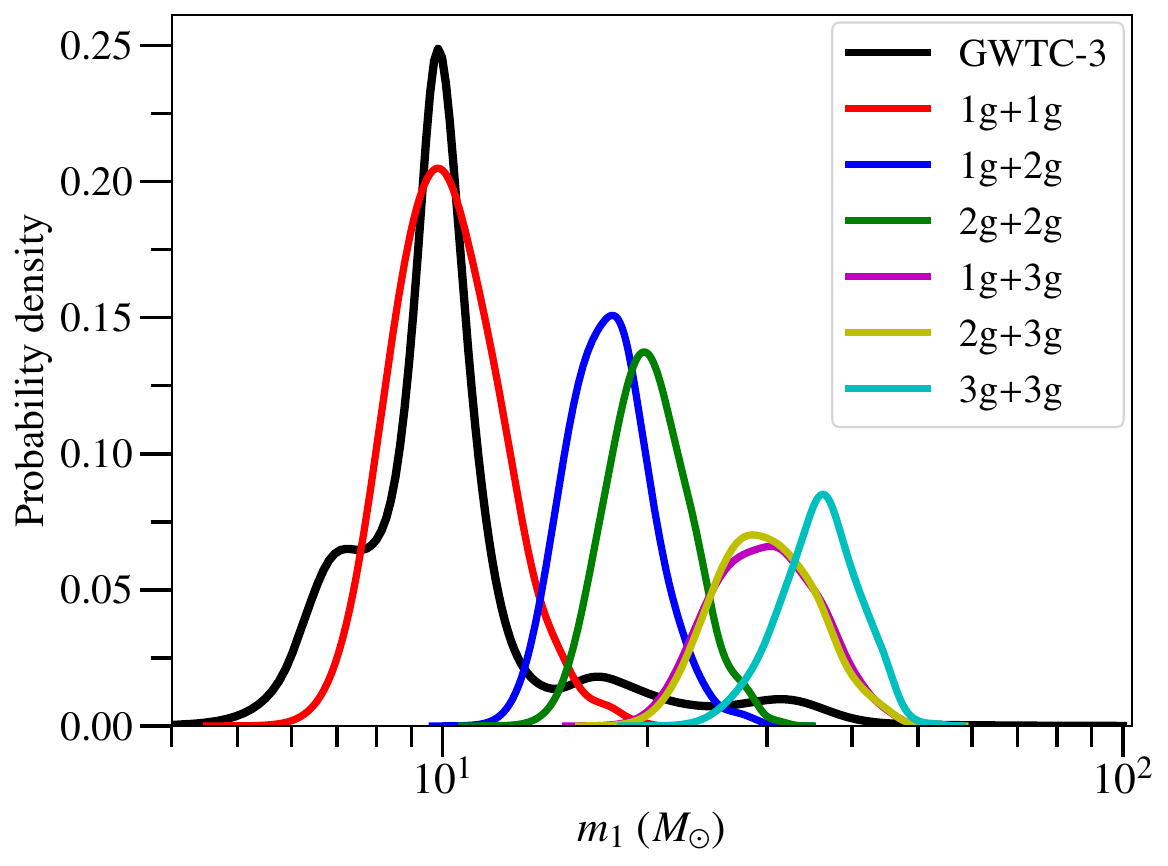}
    \caption{The mass distribution of the primary binary component. The colored curves show the different generations of BBH mergers computed via our {\tt SPHM} with $\alpha=8$, $\delta_{m}=10M_\odot$, $m_{\rm min}=5M_\odot$, and $m_{\rm max}=20M_\odot$, and $(\beta_1=4,\, \beta_2=3)$ for the pairing exponents in Eq.~(\ref{eq:Pair-q-Mtot}). The black curve shows the observed GWTC-3 population fitted by the FM model (see text).
    The escape speed of the cluster is assumed to be $400\, {\rm km/s}$ in our model; we have verified that this assumption has no visible impact on the model predictions for $V_{\rm esc}\gtrsim200$ km/s.
    Note that the mass distributions are normalized individually. Therefore the relative heights between them carry no information.
    }
    \label{fig:Pm1}
\end{figure}

\begin{table}
 \caption{\label{tab:mass-spectrum} 
 {Fraction of each binary merger type falling near the peaks in the primary mass distribution.
 The first column indicates the primary component mass ranges corresponding to the first, second, and third peaks in the GWTC-3 mass spectrum shown in Fig.~\ref{fig:Pm1}. The second column lists the binary merger type. The third column is the fraction of the indicated merger type that contributes to a given primary mass range.
 Numbers in bold indicate the mass range where a given merger type makes its dominant contribution. Some fractions do not add up to $100\%$ because those merger types have non-negligible support outside of the indicated mass ranges.
 } 
}
\begin{center}
\begin{tabular}{|c|c|c|}
\hline 
Primary mass range  & Merger type & Fraction\\
\hline \hline
 \multirow{6}{*}{$5 \rm{M_{\odot}} - 13 \rm{M_{\odot}}$} & 1g+1g & \textbf{0.87}\\
   &  1g+2g & 0.00 \\
   &  2g+2g & 0.00\\
   &  1g+3g & 0.00\\
   &  2g+3g & 0.00\\
    &  3g+3g & 0.00\\
\hline
 \multirow{6}{*}{$13 \rm{M_{\odot}} - 25 \rm{M_{\odot}}$} & 1g+1g & 0.13\\
   &  1g+2g & \textbf{0.98} \\
   &  2g+2g & \textbf{0.92}\\
   &  1g+3g & 0.13\\
   &  2g+3g & 0.11\\
    &  3g+3g & 0.00\\
\hline
 \multirow{6}{*}{$25 \rm{M_{\odot}} - 44 \rm{M_{\odot}}$} & 1g+1g & 0.00\\
   &  1g+2g & 0.02\\
   &  2g+2g & 0.08\\
   &  1g+3g & \textbf{0.86}\\
   &  2g+3g & \textbf{0.87}\\
    &  3g+3g & \textbf{0.92}\\
\hline
\end{tabular}
\end{center}
\end{table}

\section{Predicted mass distribution}\label{sec:Pm1}
The {\tt SPHM} predicts the mass spectrum of various BH generations as a function of the model parameters ($\alpha, \, \delta_{m}, \, m_{\rm min}, \, m_{\rm max}$)
and choice of pairing probability function. This allows us to compare model predictions from the {\tt SPHM} with the GWTC-3 results. Currently, two types of phenomenological population models---parametric models and nonparametric models---are applied to the GW data to understand the properties of the BBH population.  Parametric population models have specific functional forms based on certain astrophysical motivations.  On the other hand, the nonparametric population models (e.g., the Flexible Mixture/FM model, \citealt{Tiwari:2021yvr, Tiwari:2020vym}, and the Power Law + Spline/PS model, \citealt{Edelman:2021zkw}) are constructed with fitting functions containing {\it considerably} more flexibility so that they can capture certain astrophysical formation scenarios. Those models generically try to fit the structures present in the data rather than perform a parametrized fit. When applied to the GW data, the nonparametric population models have revealed multimodal substructures in the BBH mass spectrum; see, for e.g., Fig.~11 and Sec.~VI of \citealt{GWTC-3-pop}, which shows the mass spectrum of the primary component of the BH binaries in GWTC-3 inferred by different models. For convenience, we focus on the FM model discussed there. 

The FM model fitted to GWTC-3 events predicts multiple peaks in the component mass spectrum.  This was first identified in the GWTC-2 population \citep{GWTC-2-pop} by \citet{Tiwari:2020otp}. The dominant peak occurs at $\sim 9 M_{\odot}$, with two subdominant peaks at $\sim 17 M_{\odot}$ and $\sim 32 M_{\odot}$ (see Fig.~11 of \citealt{GWTC-3-pop}). \citet{Tiwari:2020otp} speculated that these multimodal features could be imprints of the hierarchical merger scenario in dense stellar environments. Here, we computed the mass spectrum of higher generation mergers from the {\tt SPHM} with the following values of the model parameters: $\alpha=8$, $\delta_{m}=10M_\odot$, $m_{\rm min}=5M_\odot$, and $m_{\rm max}=20M_\odot$, and $(\beta_1=4,\, \beta_2=3)$.
We considered the pairing probability function in Eq.~(\ref{eq:Pair-q-Mtot}).
In Fig.~\ref{fig:Pm1} we compare our predictions of the primary mass spectrum to those of the FM model (solid black).
Intriguingly, the dominant peak is consistent with the 1g+1g mergers of the {\tt SPHM}. The secondary and tertiary peaks are consistent with mergers involving 2g (1g+2g and 2g+2g mergers) and 3g (1g+3g, 2g+3g, and 3g+3g mergers) BHs, respectively. These curves assume a uniform distribution between [0, 0.2] for spin magnitudes; but we have verified that using a beta distribution for the spin magnitudes does not produce any visible change in the plots.

Note that the above adopted parameters for the BH initial mass function resembles the population present in clusters with higher metallicities ($0.015<Z<0.0225$; see Fig.~\ref{fig:IMF-fit}). 
Therefore, the hierarchical merger scheme could potentially explain the multimodality in the GWTC-3 mass spectrum if the observed BBHs primarily originate from high metallicity clusters under the assumptions of this study. We also find similar features when we compare the chirp mass distribution of the GWTC-3 population (Fig.~2 of \citealt{GWTC-3-pop}) with the chirp mass distribution of higher generation mergers as predicted by the {\tt SPHM}.

The multimodal features in the mass spectrum of the BBH population---unveiled by the nonparametric population models---can therefore be interpreted by invoking the hierarchical merger scenario in dense high-metallicity stellar environments. The occurrence of multiple peaks in the mass spectrum is attributed to the different merger generations within the hierarchical merger scenario.
Under the assumption that the distinct peaks in the observed component mass spectrum arise from different generations of mergers, the {\tt SPHM} can be used to quantify the contribution of each merger type to the different mass ranges encompassing the three peaks in the observed mass spectrum. To do this, we consider three mass ranges that broadly capture the first, second, and third peaks: $5 M_{\odot}\mbox{--}13 M_{\odot}$, $13 M_{\odot}\mbox{--}25 M_{\odot}$, and $25 M_{\odot}\mbox{--}44 M_{\odot}$. The fractions of Mg+Ng mergers that lie inside these mass ranges are shown in Table \ref{tab:mass-spectrum}. Assuming the pairing function of Eq.~(\ref{eq:Pair-q-Mtot}), we find that $87\%$ of the 1g+1g BBHs have primary masses between $5 M_{\odot}\mbox{--}13 M_{\odot}$;  $98\%$ of the 1g+2g mergers and $92\%$ of the 2g+2g mergers have primary masses in the range $13 M_{\odot}\mbox{--}25 M_{\odot}$; $86\%$ of the 1g+3g mergers, $87\%$ of the 2g+3g mergers, and $92\%$ of the 3g+3g mergers have primary masses within the $25 M_{\odot}\mbox{--}44 M_{\odot}$ mass range. According to the {\tt SPHM}, this suggests that the first peak is dominated by 1g+1g mergers, the second is dominated by 1g+2g and 2g+2g mergers, and the third peak is dominated by 1g+3g, 2g+3g and 3g+3g mergers. For instance, the {\tt SPHM} interprets GW190412 \citep{GW190412} as a 1g+3g merger which is consistent with the findings of \citet{Rodriguez:2020viw}, where GW190412 was explained as a 1g+3g merger in massive super star clusters with high metallicities ($Z\sim0.02$) and large central escape speeds ($V_{\rm esc}\sim300$ km/s).

Despite the small sample size of the currently observed BBH population ($\sim 84$ BBH mergers), these findings strongly suggest that the multiple peaks in the observed mass spectrum could originate from different generations of mergers in high-metallicity clusters ($0.015<Z<0.0225$). With the expected detection of several hundred BBHs in future observing runs, these peaks---if real---will be much better resolved, allowing for more precise vetting of the model predictions. Additional detections will also help infer the values of the $\alpha$, $\beta_1$ and $\beta_2$ power-law exponents. These are fundamental quantities that govern the formation and pairing of BHs in dynamical formation scenarios.

\section{Conclusions and caveats}\label{sec:conclusion}
We have proposed a computationally inexpensive parametric model {\tt SPHM} to study the forward evolution of the binary black hole population in massive star clusters. The free parameters of the {\tt SPHM} are broadly related to the physics and astrophysics of star formation and binary pairing/dynamical exchange in dense stellar environments. Though the {\tt SPHM} neglects physical phenomena such as BH ejections due to multi-body interactions and the evolution of clusters, the free parameters of the model are sensitive to cluster metallicity, escape speed and mass segregation. We compared predictions from the {\tt SPHM} with the {\tt cBHBd} model for a few representative configurations and found good agreement; this suggests that our model ({\tt SPHM}) can effectively capture some of the generic features of the hierarchically formed stellar mass BBH population in massive clusters. 

Major findings of this work include: \\
\indent (i) Comparing our results with those for representative configurations from {\tt cBHBd}, we find that the mass spectrum for hierarchical mergers is well-approximated by {\tt SPHM} by tuning the various parameters of the model.
   \\
\indent (ii) The retention probabilities of BBHs in dense clusters decrease significantly as we go from first generation to higher generation mergers. However, higher generation asymmetric merger types (e.g., 1g+2g, 1g+3g, etc.) have higher retention probabilities than those of the same type (e.g., 2g+2g, 3g+3g etc.), leading to an increased likelihood of such mergers. The retention probability of merger remnants falls off from $\sim 61\%$ $(\sim 10\%)$ to $\sim 13\%$ ($\sim 1\%$) as we proceed from first generation to higher generation mergers in NSCs (GCs).
\\
\indent (iii) Finally,  by applying our model {\tt SPHM} to the BBH detections to date, we find the model parameters that fit the observed mass spectrum. These parameters suggest that, if hierarchical mergers are responsible for the multiple peaks, these mergers should be occurring in clusters with high metallicity ($0.015<Z<0.0225$). This metallicity range is significantly higher than the majority of globular clusters in our Galaxy \citep{Harris:1996}. However, the Milky Way nuclear star cluster may very well have a large fraction of stars with super-solar metallicities \citep{Do2018,Do2020,Chen2023}.

The agreement between {\tt SPHM} and {\tt cBHBd} suggests that it should be possible to develop a hierarchical Bayesian framework to constrain the properties of dynamically formed BBHs using the {\tt SPHM}. More specifically, we can treat the free parameters ($\alpha, \, \delta_{m}, \, m_{\rm min}, \, m_{\rm max}, \, \beta_1, \, \beta_2$) of the {\tt SPHM} along with branching ratios for the different merger generations (controlling the population shape) as ``hyperparameters'' of the BBH population. We can estimate these hyperparameters from the observed BBH catalog using a hierarchical Bayesian inference scheme that accounts for observational biases. The inference of these hyperparameters will help constrain (i) the mass and spin distributions of first-generation black holes, (ii) the relative merger rates of the different generations of BBH, and (iii) the properties of host astrophysical environments (like metallicity, escape speed etc.).

\subsection{Caveats and possible extensions of the model}
Despite the agreement with {\tt cBHBd} and the GWTC-3 mass spectrum, our model neglects several important physical ingredients that affect the modelling of dense clusters. The most important of these are discussed below:\\
\indent (i) A major shortcoming of the model is that it is not dynamical. That is, the model does not consider the time evolution of the binary population or the star clusters. In particular, we do not account for the delay times between BBH mergers or the time taken for a remnant BH to form a next generation binary. \\
\indent (ii) Similarly, our model does not account for the expected redshift evolution of star cluster properties. The cluster escape speed plays a particularly important role in our model. Our conclusions only apply for processes that take place on timescales shorter than the time for the escape speed to vary significantly. Hence, evolving our model beyond the 3rd or 4th generations may not produce reliable results. \\
\indent (iii) Lastly, our model does not distinguish between different types of BBH formation mechanisms that may happen in a cluster (e.g., capture, 3-body interactions, etc.) and the branching ratios between those models.  

There are natural extensions of our model that may improve agreement with $N$-body simulation results. Including the delay time distribution of the binaries is one obvious ingredient to consider. Here we focuses on the mass spectrum of hierarchically formed stellar mass BBHs. One can also include additional free parameters in our model to constrain the spin distributions of hierarchical-formed BBHs. We also assumed that all BBHs have circular orbits. One can introduce more free parameters in the model to study the eccentricity distributions of hierarchical-formed BBHs. These issues will be pursued in future work. 

As is often the case, simple models are able to phenomenologically capture effects of complex phenomena. Our model seems to be in this category.
It provides a computationally inexpensive method for carrying out statistical analyses of BBH population data. However, more rigorous comparisons of this model with $N$-body models will be required in order to reliably map the best-fit parameters to cluster properties which is planned as a follow up project.

\section*{Acknowledgements}
This material is based upon work supported by the NSF’s LIGO Laboratory, which is a major facility fully funded by the National Science Foundation (NSF). K.G.A.~acknowledges support from the Department of Science and Technology and Science and Engineering Research Board (SERB) of India via the following grants: Swarnajayanti Fellowship Grant No. DST/SJF/PSA-01/2017-18 and MATRICS grant (Mathematical Research Impact Centric Support) MTR/2020/000177. K.G.A and P.M. acknowledge the support of the Core Research Grant No. CRG/2021/004565 of the Science and Engineering Research Board of India and a grant from the Infosys foundation.  K.G.A and B.S.S. acknowledge the support of the Indo-US Science and Technology Forum through the Indo-US Centre for Gravitational-Physics and Astronomy, Grant No. IUSSTF/JC-142/2019. D.C. is supported by the STFC Grant No. ST/V005618/1. D.C. and B.S.S. are grateful to the Aspen Center for Physics (ACP) summer workshop 2022 for fostering discussions that enriched this collaborative project.
We also acknowledge NSF support via NSF awards No. AST-2205920 and No. PHY-2308887 to A.G., NSF CAREER award No. PHY-1653374 to M.F., and No. PHY-2207638, No. AST-2307147, No. PHY-2308886, and No. PHYS-2309064 to B.S.S. 
This research has made use of data obtained from the Gravitational Wave Open Science Center (www.gw-openscience.org), a service of LIGO Laboratory, the LIGO Scientific Collaboration and the Virgo Collaboration. Virgo is funded by the French Centre National de Recherche Scientifique (CNRS), the Italian Istituto Nazionale della Fisica Nucleare (INFN), and the Dutch Nikhef, with contributions by Polish and Hungarian institutes. This manuscript has the LIGO preprint number P2200265. 

\section*{DATA AVAILABILITY}
The data underlying this article will be shared on reasonable request to the corresponding authors.


\bibliography{ref-list}

\clearpage


\appendix*
\onecolumngrid
\section{Supplemental Material}
\subsection{\label{sec:step-appen} Summary of our BBH population evolution scheme}
In this Supplement, we provide additional details and a step-by-step schematic procedure illustrating how we form binary black holes, pair them, and forward evolve the binary population.

\begin{enumerate}
\item Using the 1g BH mass [see Eq.~(\ref{eq:IMF})] and spin distributions as described in Sec.~\ref{sec:model}, we draw a set of 1g BH mass values $\{m_j\}$ and spin values $\{{\bm \chi}_j\}$, where $j=1\ldots 2N$ and $N\sim 10^6$. This yields a set of unpaired 1g BHs characterized by their masses and spins $\Theta_j^{\rm 1g} = [m_j,{\bm \chi_j}]$, 
\begin{equation}{\label{eq:1gBHs}}
    \{ \Theta_j^{\rm 1g}\} = \{ \Theta_1^{\rm 1g}, \Theta_2^{\rm 1g}, \ldots, \Theta_{2N}^{\rm 1g} \} \,.
\end{equation}

\item These BHs are now randomly paired to create a list of $N$ 1g+1g BH/BH pairs with parameters
\begin{equation}{\label{eq:1gBHpairs}}
\{\theta_i^{\rm 1g+1g}\} = \{ \theta_1^{\rm 1g+1g}, \theta_2^{\rm 1g+1g}, \ldots, \theta_N^{\rm 1g+1g} \} \,,
\end{equation}
where
$\theta_i^{\rm 1g+1g} = [ m^{(i)}_1, m^{(i)}_2, {\bm \chi}_1^{(i)}, {\bm \chi}_2^{(i)} ]$ are the parameters for the $i^{\rm th}$ 1g+1g BH/BH pair. 

\item From this set of binary parameters we compute the pairing probability for each pair using Eq.~(\ref{eq:Pair-q}), (\ref{eq:Pair-Mtot}), or (\ref{eq:Pair-q-Mtot}). For illustration here, we will use the pairing in Eq.~(\ref{eq:Pair-q}). (For other pairing scenarios the steps are very similar). From the set of binary parameters we compute the mass ratio for each pair $\{q_i\} = \{q_1, q_2, \ldots, q_N\} = \{m_2^{(1)}/m_1^{(1)}, m_2^{(2)}/m_1^{(2)}, \ldots,$ $ m_2^{(N)}/m_1^{(N)} \}$. The pairing probability for each pair in the set is then constructed via Eq.~(\ref{eq:Pair-q}), $p^{\rm pair} = A q^{\beta}$:
\begin{equation}\label{eq:Ppairs}
    \{p^{\rm pair}_i\} = A \{q_1^{\beta}, q_2^{\beta}, \ldots, q_N^{\beta}  \}
\end{equation}
with
\begin{equation}\label{eq:Norm}
    A = \frac{1}{\sum_{i=1}^N q_i^{\beta}} \,.
\end{equation}
    
\item These 1g+1g BH/BH pairs are then drawn in proportion to their pairing probabilities $p^{\rm pair}$ to form 1g+1g BBHs. 
For a cluster with $N$ 1g+1g binaries, the effective number of binaries with the parameter combination $\theta_i^{\rm 1g+1g}$ is then $N p^{\rm pair}_i.$
However, in practice no BH pairs are thrown away in the pairing process, they are simply down weighted according to their $p^{\rm pair}$ value so that their appearance in our dataset is rarer.
Binning this list for a particular parameter (e.g., the primary mass) then allows one to construct histograms as a function of the 1g+1g binary parameters for a given cluster. 

\item For each of the $N$ 1g+1g binaries, we use NR fitting formulas \citep{Barausse:2012qz, Hofmann:2016yih,Lousto:2012su} to calculate the final mass $M_f$, final spin $\chi_f$, and kick magnitude $V_{\rm kick}$ of the merger remnants resulting from each {\it binary} pair:
\begin{align}
\{M_f^i \} &= \{ M_f(\theta_1^{\rm 1g+1g}), M_f(\theta_2^{\rm 1g+1g}), \ldots, M_f(\theta_N^{\rm 1g+1g}) \} \,,\\       
\{\chi_f^i \} &= \{ \chi_f(\theta_1^{\rm 1g+1g}), \chi_f(\theta_2^{\rm 1g+1g}), \ldots, \chi_f(\theta_N^{\rm 1g+1g}) \} \,, \\       
\{V_{\rm kick}^i \} &= \{ V_{\rm kick}(\theta_1^{\rm 1g+1g}), V_{\rm kick}(\theta_2^{\rm 1g+1g}), \ldots, V_{\rm kick}(\theta_N^{\rm 1g+1g}) \} \,.       
\end{align}

\item Consider $K$ clusters with escape speeds $\{  V^{\rm esc}_k \}=\{  V^{\rm esc}_1, V^{\rm esc}_1, \ldots,V^{\rm esc}_K\}$. For each cluster with value $V^{\rm esc}_k$ in the list, we compare with the kick of all binaries in $\{V_{\rm kick}^i \}$ and remove those binaries for which $V_{\rm kick}^i > V^{\rm esc}_k$. This results in a list of 2g BH progenitors with mass and spin parameters $\{ \Theta^{\rm 2g}_i \} = \{ \Theta^{\rm 2g}_1, \Theta^{\rm 2g}_2, \ldots, \Theta^{\rm 2g}_{N'} \}$, where $\Theta_i^{\rm 2g} =[M_f^i,\chi_f^i]$, $N-N'$ is the number of 1g+1g binaries that are ejected by the cluster with $V^{\rm esc}_k$, and $N'$ is the number that are retained. The retention probability for 1g+1g BBHs is given by $N'/N$, i.e., the fraction of binaries with $V_{\rm kick}^i < V^{\rm esc}_k$. 
We repeat this exercise for all values of $V^{\rm esc}_k$ up to $V^{\rm esc}_K$. Each of the $K$ clusters with a given escape speed value carries its own list of retained 2g BH remnants with parameters $\{ \Theta^{\rm 2g}_i \} = \{ \Theta^{\rm 2g}_1, \Theta^{\rm 2g}_2, \ldots, \Theta^{\rm 2g}_{N'} \}$; these represent the mass and spin distributions for the 2g progenitor BHs. Note that the spin tilt angles are always drawn isotropically over a sphere as mentioned in Sec.~\ref{sec:model}.   
\end{enumerate}    

In the next generation there are two possibilities: 1g+2g and 2g+2g binaries. To construct the 1g+2g BBH population we draw $N$ 1g+2g BH/BH pairs, $\{\theta_j^{\rm 1g+2g}\}=\{\Theta^{\rm 1g}_j, \Theta^{\rm 2g}_j$\}, where $\Theta^{\rm 1g}_j$ is drawn from the 1g BH population and $\Theta^{\rm 2g}_j$ is drawn from the 2g BH population (i.e., from the population of retained 1g+1g remnants, $\{ \Theta^{\rm 2g}_i\}$). Then we repeat the steps 3, 4, 5, and 6 to estimate the properties of 1g+2g BBHs and their remnants. We similarly estimate the properties of 2g+2g BBHs. Retained remnants from 1g+2g and 2g+2g mergers will form the 3g progenitor BH population, $\{\Theta_i^{\rm 3g}\}$. Repeating the algorithm allows us to estimate the properties of third generation binaries and their remnants.

\end{document}